\documentclass{article}

\usepackage{isolatin1}
\usepackage[english]{babel}
\usepackage{graphicx}
\usepackage{amssymb, amsmath}
\usepackage[margin=3cm]{geometry}
\renewcommand{\vec}[1]{\boldsymbol{#1}}
\newcommand{\mat}[1]{\boldsymbol{#1}}
\newcommand{\ave}[1]{\langle #1 \rangle}
\newcommand{\cave}[1]{\overline{#1}}
\newcommand{\eps}{\varepsilon}

\begin{document}

\title{De Gustibus Disputandum}

\author{
	\begin{minipage}{9cm}
	\begin{center}
	\begin{description}
	\item Franco Bagnoli$^{1,2,3,4}$ (\texttt{bagnoli@dma.unifi.it})\\ 
	\item Arturo Berrones$^1$ (\texttt{arturo@dma.unifi.it}\\
	\item Fabio Franci$^{1,2,4}$ (\texttt{fabio@dma.unifi.it}\\
	\end{description}
	\end{center}
	\end{minipage}\\[1cm]
	\small
	\begin{minipage}{12cm}
	\begin{description}
	\item[1]{Dipartimento di Energetica, Universit\`a di  Firenze, 
	Via S. Marta, 3 I-50139 Firenze, Italy}
	\item[2]{INFM, Sezione di Firenze}
	\item[3]{INFN, Sezione di Firenze}
	\item[4]{Centro interdipartimentale per lo Studio delle Dinamiche 
	Complesse, Firenze}
	\end{description}
	\end{minipage}
}

\maketitle

\begin{abstract}

We propose a simple method to predict individuals' expectations about
products using a knowledge network. As a complementary result, we
show that the method is able, under certain conditions, to extract
hidden information at neural level from a customers' choices database.
\end{abstract}

\section{Introduction}

Personal tastes are universally considered very difficult to be
analyzed (there's an old latin proverb stating ``De gustibus non
disputandum est'', i.e. ``There's no accounting for taste''),
nevertheless, there is evidence of certain regularities in personal
preferences, allowing people to successfully choose Christmas
presents for friends and relatives. 

Indeed, any modeling of market behavior assumes a rational behavior
of agents, that choose among the possible options on quantitative
basis. In a famous paper~\cite{Stigler}, Stigler and Becker argued 
that all people have fixed tastes except for small variations, and that  
the different patterns in taste investments (like buying new music
disks) are computable from the expectations in revenues (i.e.\ the
forecasted enjoy in future music exposure). Many  people have argued
that this purely economic point of view is ignoring the 
``enormous role of
historical and cultural forces, education, and values, as the initial
shapers of our preferences''~\cite{Etzioni}.

Anyhow, in order to test any economic/psychologic/moral point of view
about tastes, we need a quantitative model of tastes formation and, more
important yet, of tastes anticipation based on past experience. 

The starting point of our analysis is that the  opinion of an
agent on a given product is formed by the match between agent's set
of preferences/tastes and product's qualities. 

While many commercial studies are based on surveys about customer's
preferences, we assume that both preferences and qualities are hidden
degrees of freedom, and that only the expressed opinion is
observable. One of the goal of our study is to develop techniques
able to extract information about the hidden parts from the
correlations among agent's opinions on products.

Let us suppose  that there exists a database of agents' opinions on a
given set of products. This database can be seen as a sparse matrix,
with holes corresponding to missing opinions (say, agents that have
never been exposed to a given product).

In geometrical words, one represents agent's preferences as a vector
in an hypothetical taste space, whose dimension and base vectors are
unknown. A product is represented by a similar vector (in dual
space). Agent's opinion on a given product is given by an operation
analogous to the scalar product between preferences and properties.
Therefore, products acts like a basis, and opinions as agent's
coordinates on such a basis. However, differently from usual
geometrical problems, we do not know what the basis is, if it is
complete, etc.

As we shall recall in Section~\ref{Model}, 
S. Maslov and Y.C. Zhang have shown that it is possible, if we know
the basis of agent's preferences, to reconstruct the vectors of the
individual tastes from the knowledge of a sparsely connected network
of the overlaps (scalar products) among preferences~\cite{MZ01}. We
want to extend this result to the more usual case in which basis
information is not at our disposal, as discussed in
Section~\ref{OurModel}. 

One of the outcome of our analysis 
is the possibility of opinion
anticipation, i.e. the possibility of exploiting the correlations in
the database to forecast the missing opinions. Alternatively, we can
obtain information about the overlaps of tastes between two
individuals from the knowledge of their expressed opinions. 

What we think is our main result, is the possibility
of extracting information about the hidden degrees of freedom, and
in particular the dimensionality of hidden space, from the opinion
database. In this way customer's commercial interests
 can be used as tools of cognitive psychology. 

As we
shall discuss in Section~\ref{Bias}, the sparseness of  data and a 
bias in the
database can be included in the model. 

The results of the comparisons between the theory and
numerical simulations over randomly-generated data are presented in
Section~\ref{Results}.

Finally, in section \ref{Future}, we summarize our work and draw
some conclusions.

\section{The Model} \label{Model}

We consider a population of $M$ individuals interacting with a set of
$N$ products. We assume that each product is characterized by an
$L$-dimensional array $\vec{a}=(a^{(1)},a^{(2)},\ldots,a^{(L)})$ of
features, while each individual has the corresponding list of $L$
personal tastes on the same features
$\vec{b}=(b^{(1)},b^{(2)},\ldots,b^{(L)})$. For numerical simulations
we have chosen both $a_n^{(l)}$ and $b_m^{(l)}$ in the set
$\{-1,1\}$. 

The opinion of individual $m$ on product $n$, denoted by
$s_{m,n}$,  is defined proportional to the internal product between
$\vec{b}_m$ and $\vec{a}_n$: $s_{m,n}=\lambda(L)
\vec{b}_m\cdot{\vec{a}_n}$, where $\lambda(L)$ is a suitably chosen
normalization factor.  
In general, $\lambda(L)$ should scale as $L^{-1}$ and depend on the
ranges of $\vec{a}$ and $\vec{b}$. For our choice of hidden
parameters, we use $\lambda(L)=1/L$, 
so that $s_{m,n}$ lies in the interval $[-1,1]$.

In order to predict whether the person
$j$ will like or dislike a certain product $\vec{a}_n$,
\emph{assuming to know $\vec{a}_n$}, it is sufficient to predict the
individual tastes of person $j$, i.e.\ the vector $\vec{b}_j$.

The similarity between tastes of two individuals $i$ and $j$ is
defined by the overlap $\Omega_{ij}=\vec{b}_i\cdot\vec{b}_j$ between
the individual tastes $\vec{b}_i$ and $\vec{b}_j$. 

One can build a knowledge network among people,
using the vectors $\vec{b}_m$ as nodes and the overlaps $\Omega_{ij}$
as edges. Maslov and Zhang~\cite{MZ01} (MZ) assume that a
 fraction $p$ of
these overlaps are known.  They show that there are two important
thresholds for $p$ in order to be able to reconstruct the
missing information.

The first one is a percolation threshold, reached when the fraction
of edges $p$ is greater than $p_1=1/M-1$ where $M$ is the number of
people. This means that there must be at least one path between two
randomly chosen nodes, in order to be able to predict the second
node starting from the first one.

Since vectors $\vec{b}_n$ lie in an $L$ dimensional space, and a
single link ``kills'' only one degree of freedom, a reliable
prediction need more than one path connecting two individuals. Maslov
and Zhang show that there is a ``rigidity'' threshold $p_2$, of the
order of $2L/M$, such that for $p>p_2$ the mutual orientation of
vectors in the network are fixed, and the knowledge of preferences of
just one person is sufficient to know those of all the rest of
individuals.

\section{Extracting information from hidden quantities} \label{OurModel}

In general one does not have access to individual's preferences. 
Nor one knows the dimensionality $L$ of this space. In order to address 
this problem, let us define the opinion correlation matrix $\mat{C}$:
\begin{equation}\label{corr}
C _{i,j}=\frac{\sum_{n=1}^{N} (s_{i,n}-\cave{s}_i)(s_{j,n}-\cave{s}_j)}
{\sqrt{\sum_{n=1}^{N}(s_{i,n}-\cave{s}_i)^{2}\sum_{n=1}^{N}(s_{j,n}
-\cave{s}_j)^{2}}},
\end{equation}
where $\cave{s}_i$ is the average of the opinion matrix 
$\mat{S}$ over column $i$.
 
We show below that one can compute an accurate opinion anticipation
$\tilde{s}_{m,n}$  of a true value $s_{m,n}$ using this formula:
\begin{equation}\label{estim}
\tilde{s}_{m,n}=\frac{k}{M}\sum_{i=1}^{M} C_{m,i}s_{i,n}
\end{equation}
where $k$ is a factor that in general depends on $L$ and 
on the statistical properties of the hidden components. 
However, it will be shown that if the components of
$\vec{a}_{n}$ and $\vec{b}_{m}$ are independent random variables,
$k$ is independent of $n$ and $m$, so it
can be simply chosen in order to 
have $\tilde{s}_{m,n}$ defined over the same interval as $s_{m,n}$. 

For instance, if we define
\begin{equation}\label{stilde} 
\tilde{s}_{m,n}^{*}=\frac{1}{M}\sum_{i=1}^{M} C_{m,i}s_{i,n},
\end{equation}
then in order to keep estimations in the range $[-1,1]$,
$k=\frac{1}{\tilde{S}_{max}^{*}}$ 
where $\tilde{S}_{max}^{*}=\max |\tilde{s}_{m,n}^{*}|$. As we shall
illustrate in the following, from this estimation of $k$ we can get
information about the dimension of hidden space $L$.

We now justify the proposed formulas for the case in which the components of
$\vec{a}_{n}$, $\vec{b}_{m}$ are independent random variables 
distributed according to
\begin{equation}\label{dist} 
P(a_{n}^{(l)}, b_{m}^{(l)})=P_{n,l}(a)P_{m,l}(b).
\end{equation} 

Averages over $P(a_{n}^{(l)}, b_{m}^{(l)})$ of any function 
$h(a_{n}^{(l)}, b_{m}^{(l)})$
are  given by
\begin{equation}\label{average1} 
\ave{h}=\sum_{m,n,l}^{\infty} h(a_{n}^{(l)}, b_{m}^{(l)})
P_{n,l}(a)P_{m,l}(b).
\end{equation}
For a set of hidden components distributed according to (\ref{dist}), 
the opinions are uncorrelated in the thermodynamic limit. 
However, the idea is that the system present fluctuations
mainly because $L$ is finite, so correlations between opinions arise and can
be used to predict unknown opinions. In order to keep the algebra simple, the
discussion will be made for the case in which the variables
$a_{n}^{(l)}$ and $b_{m}^{(l)}$
have zero mean. At the end a
generalization to biased components will be given. 

The components can be written in matrix form as
\begin{equation}\label{AB}
\mat{A}=\left( \begin{array}{ccc}
a_{1,1} & ... & a_{N,1}\\
. & & .\\
. & & .\\
. & & .\\
a_{1,L} &... & a_{N,L}
\end{array}
\right)  \qquad \qquad
\mat{B}=\left( \begin{array}{ccc}
b_{1,1} & ... & b_{M,1}\\
. & & .\\
. & & .\\
. & & .\\
b_{1,L} &... & b_{M,L}
\end{array}
\right),
\end{equation}
so the opinion matrix is defined by
\begin{equation}\label{opmatrix}
\mat{S}=\lambda(L) \mat{B}^{T}\mat{A},
\end{equation}
where $\lambda(L)$ is the normalization constant. 

The opinion correlation matrix is essentially equivalent to
\begin{equation}\label{corr2}
\mat{C}=\frac{\mat{S}\mat{S}^{T}}
{N\ave{s^{2}}},
\end{equation}
where $\ave{s^{2}}$ denotes the average of $s^{2}$ over
$P_{n,l}(a)P_{m,l}(b)$. Because of the finite size of the system, 
there are differences
between the normalization factors in definitions (\ref{corr}) and 
(\ref{corr2}). These differences are small for large $N$ and $L$, and we 
neglect them at this point because they give  non dominant
contributions to
errors in the final expressions.
An element of the opinion matrix  $\mat{S}$ is expressed by the
internal product
\begin{equation}\label{opelement}
s_{m,n}=\lambda(L) \sum_{l=1}^{L} b_{m,l} a_{n,l},
\end{equation}
so averaging over the distribution $P_{n,l}(a)P_{m,l}(b)$
\begin{equation}\label{vars}
\ave{s^{2}}=\lambda^{2}(L) L\ave{a^{2}}\ave{b^{2}}.
\end{equation}
Using (\ref{vars}) the correlation matrix can be written as
\begin{equation}\label{c2}
\mat{C}=\frac{\mat{B}^{T}\mat{A}\mat{A}^{T}\mat{B}}
{LN\ave{a^{2}}\ave{b^{2}}}.
\end{equation}
Let us now consider the expression
\begin{equation}\label{antici}
\frac{1}{M}\mat{C}\mat{S}=
\frac{\lambda(L) \mat{B}^{T}\mat{A}\mat{A}^{T}\mat{B}\mat{B}^{T}\mat{A} }
{LN\ave{a^{2}}\ave{b^{2}}}.
\end{equation}
If $N$ and $M$ are large, the central
limit theorem can be applied to the following matrix products
\begin{equation}\label{AA}
\mat{A}\mat{A}^{T}=\ave{a^{2}}
\left( \begin{array}{ccc}
[N+{\mathcal O}(\sqrt{N})+...] &  
{\mathcal O}(1) & ...\\
{\mathcal O}(1) &  [N+{\mathcal O}(\sqrt{N})+...]& ...\\
. & . \\
. & . 
\end{array}
\right),
\end{equation}
\begin{equation}\label{BB}
\mat{B}\mat{B}^{T}=\ave{b^{2}}
\left( \begin{array}{ccc}
[M+{\mathcal O}(\sqrt{M})+...] &  {\mathcal O}(1) & ...\\
{\mathcal O}(1) &  [M+{\mathcal O}(\sqrt{M})+...] & ...\\
. & . \\
. & . 
\end{array}
\right).
\end{equation}
Introducing Eqs.~(\ref{AA}) and (\ref{BB}) into Eq.~(\ref{antici}) we
obtain
\begin{equation}\label{antici2}
\frac{1}{M}\mat{C}\mat{S}
=\frac{\mat{S}}{L}
+
{\mathcal O}\left(\left[\frac{L-1}{L}\right]
\left[
\frac{1}{\sqrt{M}}+\frac{1}{\sqrt{N}}\right]
+\frac{1}{L\sqrt{MN}}+... \right)
\mat{S}.
\end{equation}
For large values of $N$ and $M$, by comparing  Eq.~(\ref{antici2}) 
with Eq.~(\ref{estim}), we can identify  the factor $k$  with
the number of components $L$, and obtain an estimate for the average
prediction error
\begin{equation}\label{error}
	 \eps = \sqrt{\frac{1}{MN}\sum_{m,n} \left(\tilde{s}_{m,n} -
	{s}_{m,n}\right)^2} \simeq \gamma L^{3/2}
	\frac{\sqrt{M}+\sqrt{N}}{\sqrt{MN}},
\end{equation}
where
\begin{equation}\label{gamma}
\gamma = \lambda(L) \sqrt{\ave{a^{2}}\ave{b^{2}}}.
\end{equation}
Formula~(\ref{error}) implies that
the predictive power of Eq.~(\ref{estim}) grows with $MN$ and diminishes
with $L$. This fact is a consequence of the decay of the correlations among
opinions with $L$, so that more amount of information is needed in order
to perform a prediction as $L$ grows. This condition can be compared
with the ``rigidity'' threshold $p_2$ in the MZ analysis. 

\section{Sparse and biased data}\label{Bias}

In the real world one cannot expect to have at his disposal a fully
connected opinion matrix. Indeed, one of the most important feature of
an anticipation system is its  hole-filling capability.

One can extend the previous formalism to sparse datasets by
considering the parameters $M$, $N$ as functions of the individual/product
pair  $(m,n)$
in the following way:
$M_{n}$ represents the available number of opinions over product $n$
given by any agent
and
$N_{m}$ is the number of opinions expressed by agent $m$ about any product.
Using formula (\ref{estim}) with the redefined parameters $M_{n}$ and $N_{m}$,
it follows from Eq.~(\ref{antici2})
that an unknown opinion $s_{m,n}$ 
can be estimated with an accuracy that scales as
\begin{equation}\label{distance}
|\tilde{s}_{m,n}-s_{m,n}|
\sim
\gamma L^{\frac{3}{2}}\left(\frac{\sqrt{M_{n}}+\sqrt{N_{m}}}
{\sqrt{M_{n}N_{m}}}
\right)
\end{equation}
for large values of $N_{m}$, $M_{n}$ and $L$.

The accuracy  of our approach can be related with
the ``rigidity'' threshold $p_2$. To illustrate this
let us consider a situation in which $N_{m}=M_{n}$ and $N=M$.
From formula (\ref{distance}) it turns out that
the relative error in the estimation of an opinion will be
\begin{equation}\label{distance2}
\frac{|\tilde{s}_{m,n}-s_{m,n}|}{|s_{m,n}|}
\sim
\frac{2L}{\sqrt{M_{n}}},
\end{equation}
so in order to have relative errors order one or less, the inequality
$M_{n} \gtrsim 4L^{2}$ must hold. This implies for
the density of known opinions
among all the elements of the opinion matrix that
\begin{equation}\label{density}
p=
\frac{\sum_{n=1}^{M}M_{n}}{M^{2}}
\gtrsim 2Lp_2,
\end{equation}
which means that our formulas work above the ``rigidity'' threshold
$p_2$.

Our formalism is generalizable to systems with biased components,
exploiting essentially the same arguments used to justify
Eq.~(\ref{estim}).
It is found that in this case the factor $k$ that appears in the
estimation formula (\ref{estim}) is given by
\begin{equation}\label{bias}
k=\left[\frac{1}{L}+\left(\frac{L-1}{L}\right)
\frac{\ave{b}^{2}}{\ave{b^{2}}}\right]^{-1}.
\end{equation}
Notice that $k$ does not depend on the $a_{n}^{(l)}$ variables, no matter if
these variables are biased or not.

The existence of a constant value of $k$ independently of $n$ and $m$ justifies the
previously proposed normalization approach $k=\frac{1}{\tilde{S}_{max}^{*}}$.
Moreover, $k$ can  be interpreted as the {\it effective} number of
components of the vector of internal preferences $\vec{b}_m$. 
For instance, if the variance $\ave{b^{2}}-\ave{b}^{2}$ is zero, then
$b_{m}^{(l)}$ can take a unique value, 
so $\vec{b}_m$ has only one effective
degree of freedom, which is reflected by the value $k=1$. 
On the other hand the variance of
$b_{m}^{(l)}$ is maximum when $\ave{b}=0$, implying the value
$k=L$ when  all the $L$
degrees of freedom are relevant.

The behavior of the distance between the anticipated and actual values
of opinions in the biased case
is again given as in Eqs.~(\ref{error}) and (\ref{distance}), with 
\begin{equation}\label{gamma2}
\gamma = \lambda \sqrt{\ave{a^{2}}[\ave{b^{2}}-\ave{b}^{2}]}
\end{equation}

The asymmetry of formulas (\ref{bias}) and (\ref{gamma2}) with respect to 
variables $a_{n}^{(l)}$ and $b_{m}^{(l)}$ 
is related to the fact that
the opinion correlation matrix $\mat{C}$ basically reflects
the overlap between the preferences of agents. To see this let us consider
the following normalized overlap
between $\vec{b}_i$ and $\vec{b}_j$
\begin{equation}\label{omega}
\Omega _{i,j}=\frac{\sum_{l=1}^{L} b_{i,l}b_{j,l}}
{\sqrt{\sum_{l=1}^{L}b_{i,l}^{2}
\sum_{l=1}^{L}b_{j,l}^{2}}}.
\end{equation}
For a large system size the opinion correlation matrix is written
\begin{equation}\label{c3}
\mat{C}=\frac{(\lambda(L)\mat{B}^{T}\mat{A}-\ave{s}\mat{1})
(\lambda(L)\mat{A}^{T}\mat{B}-\ave{s}\mat{1})}
{N[\ave{s^{2}}-\ave{s}^{2}]}.
\end{equation}
By introducing the product $\mat{A}\mat{A}^{T}$ given in Eq.~(\ref{AA}) on 
formula (\ref{c3}), it is found that 
\begin{equation}\label{tt}
C_{i,j}= 
\Omega_{i,j}\left[1
+{\mathcal O}\left(\frac{1}{\sqrt{N}}\left(1+\frac{1}{\sqrt{L}}\right)
\right)\right],
\end{equation}
and the average error
\begin{equation}\label{sigma}
\sigma = \frac{1}{M}\sqrt{\sum_{m,m'}
\left|C_{m,m'}-\Omega_{m,m'}\right|}
\end{equation}
should grow like $\sigma \sim N^{-1/2}$.

Eq.~(\ref{tt}) states that for increasing $N$  
the correlation between
the expressed opinions of agents $i$ and $j$ tends to be
equivalent to the overlap $\Omega_{i,j}$.

\section{Numerical Results} \label{Results}
\begin{figure}
\centerline{\includegraphics[width=10cm]{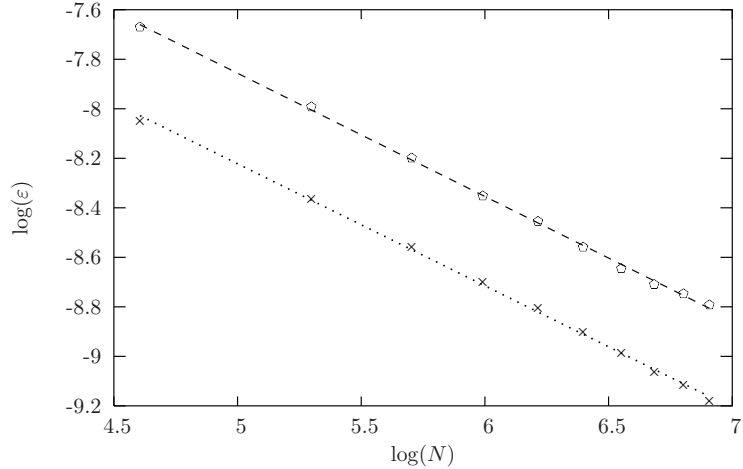}}
\caption[]{\small\label{fig1}
Average estimation error $\eps$ for $L=10$ 
as a function of number of products
$N$ for two values of $M$: $M=500$ (circles) and $M=1000$ (crosses).
The lines represent the best linear fit, with exponent $-0.498$ and
$-0.493$, resp.}
\end{figure}

\begin{figure}
\centerline{\includegraphics[width=10cm]{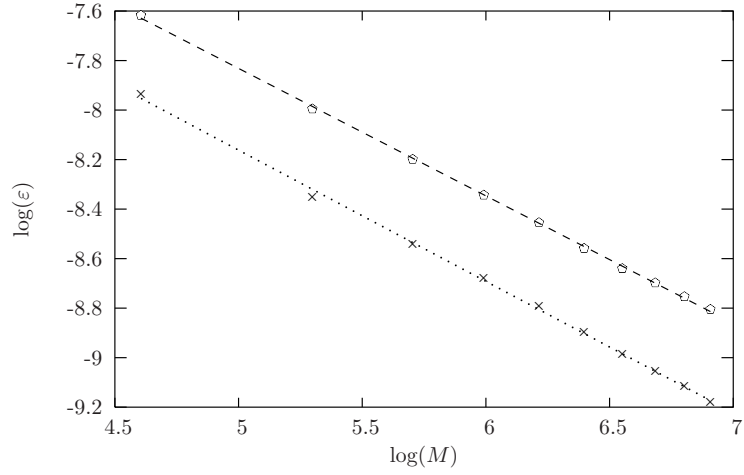}}
\caption[]{\small\label{fig2}
Average estimation error $\eps$ for $L=10$  as a function of population size
$M$ for two values of $N$: $N=500$ (circles) and $N=1000$ (crosses).
The lines represent the best linear fit, with exponent $-0.515$ and
$-0.530$, resp.}
\end{figure}

\begin{figure}
\centerline{\includegraphics[width=10cm]{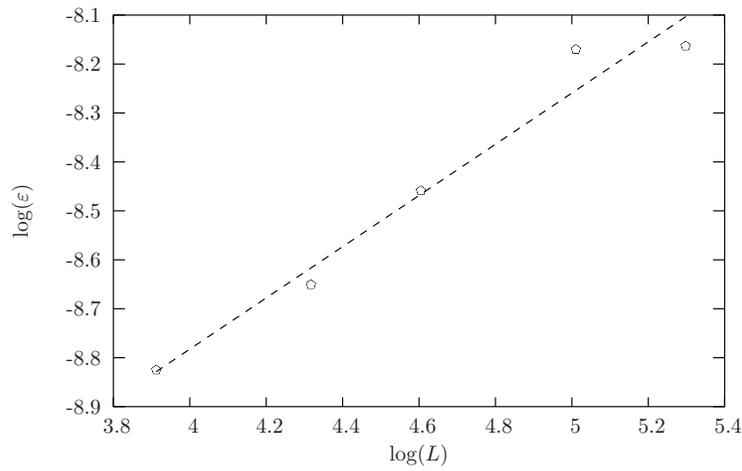}}
\caption[]{\small\label{fig3}
Average estimation error $\eps$ as a function of 
$L$ for $N=M=400$.
The line represent the best linear fit, with exponent $0.523$.}
\end{figure}

In order to test the obtained relationships, we have performed simple  
simulations using random data. 

The quantities $L$, $M$ and $N$ are free parameters.  We
have used discrete components in the $\{-1,1\}$ set, randomly 
generated with variable average. 

We have computed the opinion matrix $\mat{S}$
(Eq.~(\ref{opmatrix})), the correlation matrix 
$\mat{C}$ (Eq.~(\ref{corr})) and 
the actual overlap matrix
$\mat{\Omega}$ (Eq.~(\ref{omega})).

Then we have iterated over all the individuals' opinions $s_{m,n}$
computing $\tilde{s}_{m,n}$ from Eq.~(\ref{estim}), accumulating the 
 average quadratic estimation error $\eps$,
Eq.~(\ref{error}). 

Figures \ref{fig1}, \ref{fig2} and \ref{fig3} show that the
theoretical average errors, Eq.~(\ref{error}), 
are in good agreement with simulations.

\begin{figure}
\centerline{\includegraphics[width=10cm]{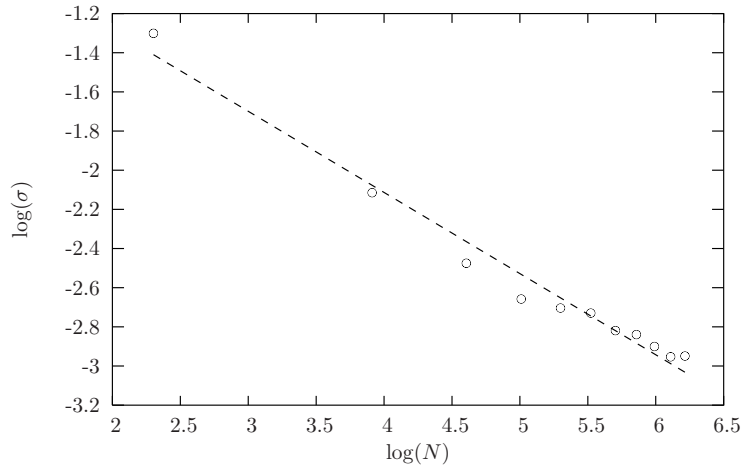}}
\caption{\label{C-Omega} Average error $\sigma$
(Eq.~(\protect\ref{sigma})) as a function of the product number $N$.
The dashed line marks the linear fitting $\sigma \sim N^{-0.41}$.   }
\end{figure}

Moreover, we show in Figure~\ref{C-Omega}
that the distance $\sigma$ between $C$ and $\Omega$ goes like
$N^{-1/2}$, as expected from Eq.~(\ref{tt}).

\section{Discussion and conclusions}\label{Future}

We assumed that an opinion is formed as a scalar product between 
individual preferences and products' properties (both
unobservable). This assumption relies on a kind of
``universality'' in cognitive processes, so that the opinion formation
process should be analogous to other brain activity like the olfactory 
system, but honestly we do not have any rigorous justification.
Individuals' opinions are assumed to be stored in a database. 

We have shown that, using central limit theorem (i.e.\ uncorrelated data) 
it is possible to anticipate an individual's  opinion, 
i.e. there is the possibility of exploiting the correlations in
the database to forecast the missing opinions. Alternatively, we can
obtain information about the overlaps of tastes between two
individuals from the knowledge of their expressed opinions. 

We have also shown that one can extract information about the
dimensionality of the hidden taste space from the opinion
database. We have also recovered the (almost trivial) expectation
 that the prediction error decreases 
when both the size of individual and product pools grow, and increases
with the dimension of the hidden space.

We have not considered here the problem of 
coevolution of tastes and product qualities (which are produced
in accordance to expectations about clients' expectations).
The
coevolution of products' features and individuals' preferences induces 
correlations: people are not expected to blindly
choose one movie from the available ones, but they tend to watch
movies based on their anticipated opinion, thus filling the dataset with
correlated data. On the other hand, movies are produced based on
market expectation, reducing still more the variability. 

The role of education emerges from this simple model: 
reliable opinion anticipations, that constitute an expectation of
``revenues'' from cultural investments, can come only from an assorted
background of experiences both from a personal point of view, but also
from the community's one (due to the need of individuals's
correlations).

Finally, this  model illustrate the value contained in 
personal information and the need for their protection. 

Experimental verifications of the model are difficult, since personal
data are jealously conserved. However, it is possible 
to identify similar ``scalar product-like'' mechanism in chemical
or biological interactions~\cite{MS02}, 
for which experimental data may be more
easily available. 

\section*{Acknowledgemnts}
We acknowledge fruitful discussions with P. Palmerini and the DOCS
group~\cite{DOCS}. A.~B. acknowledges financial support by CONACYT.

\end{document}